\documentclass[aps,pre,twocolumn,groupedaddress]{revtex4-2}
\usepackage{amsmath,amssymb,bm,derivative,graphicx,here,physics2}

\usephysicsmodule{ab}

\begin{document}

\title{Lower bound of entropy production in an underdamped Langevin system with normal distributions}
\author{Futa Watabe and Koji Okuda}
\affiliation{Department of Physics, Hokkaido University, Sapporo 060-0810, Japan}
\date{\today}

\begin{abstract}
	We study the lower bound of the entropy production in a one-dimensional underdamped Langevin system constrained by a time-dependent parabolic potential.
	We focus on minimizing the entropy production during transitions from a given initial distribution to a given final distribution taking a given finite time.
	We derive the conditions for achieving the minimum entropy production for the processes with normal distributions, using the evolution equations of the mean and covariance matrix to determine the optimal control protocols for stiffness and center of the potential.
	Our findings reveal that not all covariance matrices can be given as the initial and final conditions due to the limitations of the control protocol.
	This study extends existing knowledge of the overdamped systems to the underdamped systems.
\end{abstract}

\maketitle

\section{Introduction}

Microscopic particles in a fluid environment undergo random motion due to thermal fluctuations, known as Brownian motion.
This phenomenon can be explained by the Langevin equation, which is described by the forces acting on the particle,
including external forces, drag forces, and random thermal forces from the environment
\cite{Uhlenbeck1930,gardiner1985handbook,risken1996fokker,sekimoto2010stochastic,Seifert_2012}.
The Langevin equation for a particle with mass $m$
and position $\bm x(t)$
is given by
\begin{align}
	\label{eq:Langevin_eq}
	m\ddot{\bm{x}}(t)
	=\bm{F}(\bm{x}(t),t)-\gamma\dot{\bm{x}}(t)+\sqrt{2\gamma T}\bm{\xi}(t).
\end{align}
Here, $\bm{F}$ represents the external force, $\gamma$ is the drag coefficient, $T$ is the temperature of the environment,
and $\bm{\xi}(t)$ denotes
the Gaussian white noise with the mean $\ab<\bm\xi(t)>=\bm0$
and the variance $\ab<\xi_i(t)\xi_j(t')>=\delta_{ij}\delta(t-t')$.
When the inertial term $m\ddot{\bm x}$ can be neglected, the system is referred to as overdamped,
while retaining this term results in an underdamped system.

By interpreting the work done by the environment
on the particle as the heat $\Delta Q$ absorbed by the system,
thermodynamic quantities in this system can be treated
\cite{sekimoto2010stochastic,Seifert_2012}.
In this context,
it has been studied what the state transition improving
the energy efficiency is and a few topics related to it have also
been considered
\cite{Schmiedl_optimal,Schmiedl_2008,PhysRevLett.108.190602,Miura2021,Miura2022,li2025decompositionmetrictensorthermodynamic}.

In recent years,
particular attention has been devoted
to the entropy production as a key quantity.
The entropy production $\Sigma$ is defined
as the sum of the entropy change of the system $\Delta S$
and the entropy change in the environment $-\Delta Q/T$:
\begin{align}
	\label{eq:def of entropy production}
	\Sigma
	:=\Delta S-\frac{\Delta Q}{T}.
\end{align}
The entropy production symbolizes the overall dissipation
in the system and the environment.
Thus,
the transitions with minimal entropy production are often desirable.

Since the state of a Langevin system at time $t$ is represented
by a probability distribution $p(t)$,
we are led to the question of what kind of transition
from time 0 to $\tau$,
with a fixed initial distribution $p(0)=p^\mathrm{ini}$
and a fixed final distribution $p(\tau)=p^\mathrm{fin}$,
minimizes the entropy production $\Sigma$
when the external force $\bm F$ is treated as a control parameter.

The previous studies have established the evolution equations
for the external force that must be satisfied
in overdamped systems \cite{dechant2019thermodynamicinterpretationwassersteindistance,nakazato2021}.
Specifically, when the distribution is restricted to the normal distribution,
it has been shown that a process in which the expectation
of the position and the square root of the variance matrix evolve
at a constant rate achieves the lower bound of the entropy production,
and it has explicitly been shown
how the external forces should be manipulated to achieve this.
The previous study \cite{dechant2019thermodynamicinterpretationwassersteindistance}
has also derived the lower bound of the entropy production
in the underdamped systems,
however,
this lower bound is not necessarily achievable
when the initial and final distributions are fixed.

In addition, to the best of our knowledge,
only a limited number of studies
on the underdamped systems
have focused on the achievable entropy production bound and the protocols required to realize it \cite{Sanders_2024},
although the studies related to the lower bound of the entropy production
in underdamped systems are active
\cite{
	PhysRevE.100.032130,
	dechant2022boundsprecisioncurrentsunderdamped,
	PhysRevE.107.014604,
	lyu2024entropyproductionunderdampedlangevin}.

In this paper,
we study the conditions for achieving the lower bound of the entropy production
with fixed initial and final distributions
in a one-dimensional underdamped system.
However, we restrict the potential
energy to the parabolic potential since the general potential
turns out to be difficult to analyze in the present study.
We also discuss the methods for constructing the processes
that satisfy these conditions.
A more mathematically advanced approach to the similar problem has been
studied by Sanders et al.\cite{Sanders_2024}.

This paper is organized as follows.
In Sec.~\ref{sec:od},
we introduce an overview of overdamped systems
in the previous studies on the lower bound of the entropy production.
Next, in Sec.~III,
we describe the underdamped system studied
in this paper and address the cases of the normal distribution.
We also present the known lower bound of the entropy production
in the underdamped systems.
In Sec.~IV,
we formulate the minimization problem for the entropy production
and derive the necessary conditions for achieving the lower bound.
Section V presents the numerical simulations
that validate our theoretical results.
In Sec.~VI,
we discuss the implications of allowing noncontinuous behavior
in the control parameters,
exploring the possibility for increasing the flexibility
in setting initial and final conditions.
Finally,
we conclude with a summary of our findings
and potential directions for future research in Sec.~VII.

\section{Overdamped system}
\label{sec:od}
In this section,
we summarize the results of the previous studies
on the lower bound of the entropy production in the overdamped system,
focusing mainly on the results
in \cite{dechant2019thermodynamicinterpretationwassersteindistance,nakazato2021}.

In the overdamped system,
the inertial term $m\ddot{\bm x}$ is ignored
in Eq.~\eqref{eq:Langevin_eq},
and the Langevin equation is given by \cite{sekimoto2010stochastic}
\begin{align}
	\dot{\bm x}
	=
	\frac1\gamma\bm F+\sqrt{\frac{2T}\gamma}\bm\xi.
\end{align}

The probability distribution $p(\bm x,t)$ for the particle's position $\bm x$ at time $t$
follows the Fokker-Planck equation \cite{risken1996fokker}:
\begin{align}
	\label{eq:continuity}
	\pdv pt
	 & =
	-\bm\nabla\cdot\ab(
	\bm\nu(\bm x,t)
	p),
	\\
	\label{eq:def of local mean velocity}
	\bm\nu(\bm x,t)
	 & :=
	\frac1\gamma\bm F
	-\frac T\gamma\bm\nabla\ln p,
\end{align}
where $\bm\nu$ denotes the local mean velocity.
Then, the heat $\odv Qt$ that the system receives per unit time
and the entropy $S$ of the system are expressed as \cite{sekimoto2010stochastic,Seifert_2012}
\begin{align}
	\label{eq:dQdt_od}
	\odv Qt & =\ab<-\bm F\cdot\bm\nu>, \\
	\label{eq:S_od}
	S       & =\ab<-\ln p>,
\end{align}
where $\ab<\cdot>$ represents the expectation value over the distribution.
From Eqs.~\eqref{eq:def of entropy production},
\eqref{eq:dQdt_od},
and \eqref{eq:S_od},
the entropy production rate $\sigma$,
which is the time derivative of the entropy production $\Sigma$,
can be derived as \cite{Seifert_2012}
\begin{align}
	\label{eq:epr od}
	\sigma(t)
	:=
	\dot\Sigma(t)
	=
	\odv St
	-\frac 1T\odv Qt
	=
	\frac\gamma T
	\ab<\ab\|\bm\nu\|^2>.
\end{align}
Thus, the entropy production from time $t=0$ to $t=\tau$
is expressed as
\begin{align}
	\Sigma(\tau)=\int_0^\tau\odif{t}\,\sigma(t)
	=
	\frac\gamma T
	\int_0^\tau\odif{t}\,
	\ab<\ab\|\bm\nu\|^2>.
\end{align}

In \cite{dechant2019thermodynamicinterpretationwassersteindistance,nakazato2021},
it is shown that the entropy production $\Sigma$
in this system has the following lower bound
when the initial and final distributions
are fixed to $p(0)=p^\mathrm{ini}$
and $p(\tau)=p^\mathrm{fin}$, respectively:
\begin{align}
	\label{eq:lower bound of entropy production}
	\Sigma(\tau)
	\geq
	\frac\gamma T
	\frac{\mathcal W(p^\mathrm{ini},p^\mathrm{fin})^2}\tau,
\end{align}
where $\mathcal W(\cdot,\cdot)$ is
the $L_2$-Wasserstein distance
between distributions defined as \cite{villani2008optimal}
\begin{align}
	\mathcal W(p,q)^2:=\inf_{\Pi\in\mathcal P(p,q)}
	\left[\int\odif{\bm x}\int\odif{\bm y}\,\|\bm x-\bm y\|^2\Pi(\bm x,\bm y)\right]^{1/2},
	\label{eq:WD}
\end{align}
where $\mathcal P$ is the set of all joint distributions of $\bm x$ and $\bm y$ whose marginals are $p(\bm x)$ and $q(\bm y)$:
\begin{widetext}
	\begin{align}
		\mathcal P(p,q)
		:= \ab\{\Pi(\bm x,\bm y):
		\int\odif{\bm y}\,\Pi(\bm x,\bm y)=p(\bm x),
		\int\odif{\bm x}\,\Pi(\bm x,\bm y)=q(\bm y)
		\}.
	\end{align}
\end{widetext}

It is known that to achieve the equality,
the distribution path $p(\cdot,t)$ must follow the geodesic
on the manifold measured by the Wasserstein distance at a constant speed \cite{dechant2019thermodynamicinterpretationwassersteindistance,nakazato2021}.

It is difficult to derive the geodesic for
general distributions $p^\mathrm{ini}$ and $p^\mathrm{fin}$,
but when both $p^\mathrm{ini}$ and $p^\mathrm{fin}$
are normal distributions the geodesic can be easily constructed, as shown below.

Here, for simplicity, we consider the system to be one-dimensional.
The Wasserstein distance between the normal distributions $p_\mathrm{A}$ and $p_\mathrm{B}$,
with means $\mu_\mathrm{A}$ and $\mu_\mathrm{B}$ and variances $\Xi_\mathrm{A}$ and $\Xi_\mathrm{B}$,
respectively,
is given by \cite{OLKIN1982257,Gelbrich1990,villani2008optimal}
\begin{align}
	\label{eq:Wasserstein for normal}
	\mathcal W(p_\mathrm{A},p_\mathrm{B})
	=
	\sqrt{
		(\mu_\mathrm{A}-\mu_\mathrm{B})^2+(\sqrt{\Xi_\mathrm{A}}-\sqrt{\Xi_\mathrm{B}})^2
	},
\end{align}
which corresponds to the two-dimensional Euclidean distance
with the mean and the square root of the variance as the axes.

Therefore,
when the initial distribution $p^\mathrm{ini}$
and the final distribution $p^\mathrm{fin}$
are given by the normal distributions
with the means $\mu^\mathrm{ini}$, $\mu^\mathrm{fin}$,
and the variances $\Xi^\mathrm{ini}$, $\Xi^\mathrm{fin}$, respectively,
the path keeping the normal distribution $p(t)$,
where the mean $\mu(t)$ and variance $\Xi(t)$ evolve as Eqs.~\eqref{eq:opt mu od} and \eqref{eq:opt xi od},
follows the geodesic at a constant speed \cite{dechant2019thermodynamicinterpretationwassersteindistance,nakazato2021}:
\begin{align}
	\label{eq:opt mu od}
	\mu(t)
	 & =
	\mu^\mathrm{ini}
	+(\mu^\mathrm{fin}-\mu^\mathrm{ini})\frac{t}{\tau},
	\\
	\label{eq:opt xi od}
	\sqrt{\Xi(t)}
	 & =
	\sqrt{\Xi^\mathrm{ini}}
	+\ab(\sqrt{\Xi^\mathrm{fin}}-\sqrt{\Xi^\mathrm{ini}})\frac{t}{\tau}
	.
\end{align}
When following this path,
the entropy production $\Sigma(\tau)$ achieves the lower bound,
which is calculated from Eqs.~\eqref{eq:lower bound of entropy production}
and \eqref{eq:Wasserstein for normal}
as
\begin{align}
	\Sigma(\tau)
	=
	\frac{\gamma}{T}\frac{
		(\mu^\mathrm{fin}-\mu^\mathrm{ini})^2
		+(\sqrt{\Xi^\mathrm{fin}}-\sqrt{\Xi^\mathrm{ini}})^2
	}{\tau}.
\end{align}
It is known that
the path Eqs.~\eqref{eq:opt mu od} and \eqref{eq:opt xi od}
is achieved when the external force $F$ is given by
\begin{widetext}
	\begin{align}
		F(x,t)
		 & =
		-k(t)\ab(x-r(t)),
		\\
		k(t)
		 & =
		\frac{T}{\ab(
			\sqrt{\Xi^\mathrm{ini}}
			+\ab(\sqrt{\Xi^\mathrm{fin}}-\sqrt{\Xi^\mathrm{ini}})\frac{t}{\tau}
			)^2}
		-\frac{\gamma}{
			\sqrt{\Xi^\mathrm{ini}}
			+\ab(\sqrt{\Xi^\mathrm{fin}}
			-\sqrt{\Xi^\mathrm{ini}})\frac{t}{\tau}
		},
		\\
		r(t)
		 & =
		\mu^\mathrm{ini}+(\mu^\mathrm{fin}-\mu^\mathrm{ini})\frac{t}{\tau}
		+\frac{\gamma(\mu^\mathrm{fin}-\mu^\mathrm{ini})}{\tau k(t)}.
	\end{align}
\end{widetext}

\section{Underdamped system}
In this section,
we introduce the underdamped Langevin system studied in this paper
and explain the known lower bound of the entropy production
in the underdamped system.

In the underdamped system,
the Langevin equation is described as
\begin{align}
	\dot{\bm x} & =\bm v,
	\\
	\dot{\bm v} & =-\frac1m\bm F(\bm x,t)
	-\frac{\gamma}{m}\bm v
	+\sqrt{\frac{2\gamma T}{m^2}}\bm\xi(t).
\end{align}

The probability distribution $p(\bm x,\bm v,t)$
for the particle's position $\bm x$ and velocity $\bm v$ at time $t$
follows the Fokker-Planck equation \cite{risken1996fokker}:
\begin{align}
	\label{eq:Fokker-Planck equation general}
	\pdv{p}{t}
	 & =
	-\bm\nabla_x\cdot(\bm vp)
	-\bm\nabla_v\cdot\ab[\ab(
		\frac1m\bm F
		-\frac\gamma m\bm v
		-\frac{\gamma T}{m^2}\bm\nabla_v\ln p
		)p].
\end{align}

Then, the heat $\odv Qt$ that the system receives per unit time
and the entropy $S$ of the system are expressed as \cite{sekimoto2010stochastic,Seifert_2012}
\begin{align}
	\label{eq:dQdt_ud}
	\odv Qt
	  & =
	\ab<\ab(-\gamma\bm v-\frac{\gamma T}m\bm\nabla_v\ln p)\cdot\bm v >,
	\\
	\label{eq:S_ud}
	S & =\ab<-\ln p>.
\end{align}
Therefore, the entropy production rate $\sigma$ is derived from
Eqs.~\eqref{eq:def of entropy production},
\eqref{eq:dQdt_ud},
and \eqref{eq:S_ud}
as \cite{Seifert_2012}
\begin{align}
	\sigma(t)
	 & =\odv St-\frac1T\odv Qt
	\\
	\label{eq:entropy production rate ud close}
	 & =	\frac{\gamma}{T}\ab<\ab\|\bm v+\frac{T}{m}\bm\nabla_v\ln p\|^2>
	\\
	\label{eq:expanded EPR}
	 &
	=
	\frac{\gamma}{T}\ab<\|\bm v\|^2>
	+\frac{\gamma T}{m^2}\ab<\ab\|\bm\nabla_v\ln p\|^2>
	+\frac{2\gamma}m\ab<\bm v\cdot\bm\nabla_v\ln p>
	\\
	\label{eq:entropy production rate ud}
	 & =	\frac{\gamma}{T}\ab<\|\bm v\|^2>
	+\frac{\gamma T}{m^2}\ab<\ab\|\bm\nabla_v\ln p\|^2>
	-\frac{2d\gamma}{m}.
\end{align}
In Eq.~\eqref{eq:expanded EPR},
$\langle \bm{v} \cdot \bm{\nabla}_v \ln p \rangle = -d$ can be calculated by integration by parts,
where $d$ is the spatial dimension of the system.

\subsection{Known lower bound of the entropy production}
We explain the lower bound of the entropy production in the underdamped system,
as shown in the previous study \cite{dechant2019thermodynamicinterpretationwassersteindistance}.

We define the marginal distribution $p_x(\bm x,t)$
and the conditional distribution $p_{v|x}(\bm x,\bm v,t)$ as
\begin{align}
	\label{eq:def merge}
	p_x(\bm x,t)           & :=\int\odif{\bm v}\,p(\bm x,\bm v,t),
	\\
	\label{eq:def conditional}
	p_{v|x}(\bm x,\bm v,t) & :=\frac{p(\bm x,\bm v,t)}{p_x(\bm x,t)}.
\end{align}
Additionally,
we define the local mean velocity $\bm{\nu}(\bm x,t)$,
which is the average velocity at the position $\bm{x}$,
as
\begin{align}
	\bm{\nu}(\bm{x}, t)
	 & :=
	\int\odif{\bm{v}}\,\bm{v} p_{v|x}(\bm{x},\bm{v}, t).
\end{align}

The evolution equation for $p_x$ can be derived
from Eqs.~\eqref{eq:Fokker-Planck equation general}
and \eqref{eq:def merge} as
\begin{align}
	\label{eq:dev of merge}
	\pdv {p_x}t
	 & =
	-\bm\nabla\cdot\ab(
	\bm\nu
	p_x).
\end{align}
From Eqs.~\eqref{eq:entropy production rate ud close} and \eqref{eq:def conditional}
\begin{align}
	\sigma(t)
	 & = \frac\gamma T\int \odif{\bm{x}} \int \odif{\bm{v}} \, \left\| \bm v+\frac{T}{m}\bm\nabla_v\ln p\right\|^2 p                              \\
	 & = \frac\gamma T\int \odif{\bm{x}} \int \odif{\bm{v}} \, \left\| \bm v+\frac{T}{m}\bm\nabla_v\ln \ab(p_{v|x} p_x)\right\|^2 p_{v|x} p_x     \\
	\label{eq:mid of ineq sigma}
	 & = \frac\gamma T\int \odif{\bm{x}} \, p_x \int \odif{\bm{v}} \, \left\| \bm v+\frac{T}{m}\bm\nabla_v\ln p_{v|x} \right\|^2 p_{v|x}          \\
	\label{eq:ineq sigma}
	 & \geq \frac\gamma T\int \odif{\bm{x}} \, p_x \left\| \int \odif{\bm{v}} \, \ab(\bm v+\frac{T}{m}\bm\nabla_v\ln p_{v|x}) p_{v|x} \right\|^2,
\end{align}
where we use
$\bm v$ independence of $p_x$
and the Cauchy-Schwarz inequality.
The integral in the norm in Eq.~\eqref{eq:ineq sigma} is simplified as
\begin{align}
	 & \qquad
	\int \odif{\bm{v}} \, \ab(\bm v+\frac{T}{m}\bm\nabla_v\ln p) p_{v|x}
	\\
	 & = \int \odif{\bm{v}} \, \bm{v} p_{v|x}
	+ \frac Tm\int \odif{\bm{v}} \, \bm{\nabla}_v p_{v|x}
	\\
	\label{eq:simplified integral}
	 & = \bm{\nu}.
\end{align}
From Eqs.~\eqref{eq:ineq sigma} and \eqref{eq:simplified integral},
we have
\begin{align}
	\label{eq:ineq cg}
	\sigma
	 & \geq
	\frac{\gamma}{T}\ab<\ab\|\bm\nu\|^2>_x,
\end{align}
where $\ab<\cdot>_x$ denotes
the expectation value taken
with respect to the marginal distribution $p_x$.
To satisfy the equality in Eq.~\eqref{eq:ineq sigma},
$\ab(\bm{v} + (T/m)\bm\nabla_v\ln p_{v|x})$ must be independent of $\bm v$,
which can be rewritten as
\begin{align}
	\label{eq:condition for equality}
	p_{v|x} = A \exp\left(-\frac{1}{2} \frac{m}{T} \|\bm{v} - \bm{c}\|^2\right),
\end{align}
using arbitrary functions $A(\bm x)$ and $\bm c(\bm x)$.

Comparing these results with those in Sec.~\ref{sec:od},
it can be seen that the evolution equation
for the marginal distribution Eq.~\eqref{eq:dev of merge}
has the same form as that for the distribution
in the overdamped system Eq.~\eqref{eq:continuity},
and the lower bound
of the entropy production rate Eq.~\eqref{eq:ineq cg}
has the same form as that
in the overdamped system Eq.~\eqref{eq:epr od}.
Therefore, repeating the discussion
of the lower bound of the entropy production
in the overdamped system in Sec.~\ref{sec:od},
we can show that the entropy production $\Sigma$
in the underdamped system satisfies
\begin{align}
	\label{eq:known}
	\Sigma(\tau)\geq
	\frac{\gamma}{T}\frac{\mathcal W(p_x(\cdot,0), p_x(\cdot,\tau))^2}{\tau}.
\end{align}
However,
to achieve the lower bound in Eq.~\eqref{eq:known},
the marginal distribution $p_x(\cdot,t)$ must follow the geodesic,
and $p_{v|x}$ must satisfy Eq.~\eqref{eq:condition for equality}
at each time $t\in [0,\tau]$.
Therefore,
the solutions satisfying the lower bound in Eq.~\eqref{eq:known} do not necessarily exist
when we specify the initial distribution $p^\mathrm{ini}$
and the final distribution $p^\mathrm{fin}$, which may not meet the
above $p_x$ and $p_{v|x}$.

\subsection{Normal distribution}
\label{subsec:nd}

Since general distributions are difficult to handle
in the underdamped system,
we restrict the distribution to the normal distribution in this paper.
It is known that the normal distribution is preserved
when the external force is derived from the parabolic potential
\cite{gardiner1985handbook},
which we assume in this paper.
Furthermore,
we consider a one-dimensional system for simplicity.

Therefore,
we define
the external force $F(x,t)$
as
\begin{align}
	F(x,t)=-k(t)(x-r(t)),
\end{align}
where $k$ and $r$ are the stiffness and the center
of the parabolic potential, respectively,
which we regard as the control parameters.

In this case, the Langevin equation is described as
\begin{align}
	\dot{x} & =v,                      \\
	\dot{v} & =-\frac{k(t)}{m}(x-r(t))
	-\frac{\gamma}{m}v
	+\sqrt{\frac{2\gamma T}{m^2}}\xi,
\end{align}
and the Fokker-Planck equation is described as
\begin{align}
	\pdv{p}{t}
	 & =
	-v\pdv{p}{x}
	+\frac{k}{m}(x-r)\pdv{p}{v}
	+\frac{\gamma}{m}\pdv{}{v}(vp)
	+\frac{\gamma T}{m^2}\pdv[order=2]{p}{v}.
	\label{eq:Fokker-Planck equation}
\end{align}

A multivariate normal distribution for $x$ and $v$, as expressed below,
serves as a solution to Eq.~\eqref{eq:Fokker-Planck equation}:
\begin{align}
	\label{eq:definition or normal distribution}
	p(\bm q,t)
	 & =
	\frac{1}{2\pi\sqrt{\det\Xi}}
	\exp\ab(
	-\frac{1}{2}(\bm{q}-\bm{\mu})^\top\Xi^{-1}(\bm{q}-\bm{\mu})
	),
	\\
	\bm{q}
	 & :=
	\begin{pmatrix}
		x \\v
	\end{pmatrix},
	\\
	\bm{\mu}(t)
	 & :=
	\begin{pmatrix}
		\mu_x \\\mu_v
	\end{pmatrix}
	=
	\ab<\bm{q}>,
	\\
	\label{eq:def of Xi}
	\Xi(t)
	 & =
	\begin{pmatrix}
		\Xi_{xx} & \Xi_{xv} \\
		\Xi_{vx} & \Xi_{vv}
	\end{pmatrix}
	:=
	\ab<(\bm{q}-\bm{\mu})(\bm{q}-\bm{\mu})^\top>,
\end{align}
where $\bm\mu$ is the mean of the variables and $\Xi$ is the covariance matrix.

From Eqs.~\eqref{eq:Fokker-Planck equation}-\eqref{eq:def of Xi},
we can obtain the following evolution
equations:
\begin{subequations}\label{eq:dynamics}
	\begin{align}
		\label{eq:mu_x}
		{\dot\mu}_x
		 & =\mu_v,
		\\
		\label{eq:mu_v}
		{\dot\mu}_v
		 & =-\frac{k}{m}(\mu_x-r)-\frac{\gamma}{m}\mu_v,
		\\
		\label{eq:xi_xx}
		{\dot\Xi}_{xx}
		 & =2\Xi_{xv},
		\\
		\label{eq:xi_xv}
		{\dot\Xi}_{xv}
		 & =
		-\frac{k}{m}\Xi_{xx}
		-\frac{\gamma}{m}\Xi_{xv}
		+\Xi_{vv},
		\\
		\label{eq:xi_vv}
		{\dot\Xi}_{vv}
		 & =
		-\frac{2k}{m}\Xi_{xv}
		-\frac{2\gamma}{m}\Xi_{vv}
		+\frac{2\gamma T}{m^2}.
	\end{align}
\end{subequations}

From Eqs.~\eqref{eq:entropy production rate ud}
and \eqref{eq:definition or normal distribution}
we can express the entropy production rate as
\begin{align}
	\label{eq:entropy production rate}
	\sigma(t)
	=\frac{\gamma}{m}\ab(\frac{m}{T}\mu_v^2+\frac{m}{T}\Xi_{vv}+\frac{T}{m}\frac{\Xi_{xx}}{\det\Xi}-2),
\end{align}
and the entropy production as
\begin{align}
	\label{eq:entropy production with normal}
	\Sigma(\tau)
	=
	\frac{\gamma}{m}\int_0^\tau\odif{t}\,\ab(
	\frac{m}{T}\mu_v^2+\frac{m}{T}\Xi_{vv}+\frac{T}{m}\frac{\Xi_{xx}}{\det\Xi}-2
	).
\end{align}

The known lower bound Eq.~\eqref{eq:known} can be
calculated from Eqs.~\eqref{eq:Wasserstein for normal},
\eqref{eq:def merge},
and \eqref{eq:definition or normal distribution} as
\begin{align}
	\label{eq:known lower bound concretely}
	\frac{\gamma}{T}\frac{
		(\mu_x(\tau)-\mu_x(0))^2
		+(\sqrt{\Xi_{xx}(\tau)}-\sqrt{\Xi_{xx}(0)})^2
	}{\tau}.
\end{align}

However, in the present case,
we can derive Eq.~\eqref{eq:known lower bound concretely}
more directly without using the Wasserstein distance,
as shown below.
For the first term of the entropy production
Eq.~\eqref{eq:entropy production with normal},
we obtain
\begin{align}
	\int_0^\tau\odif{t}\,\mu_v^2
	 & =
	\label{eq:temp 1}
	\int_0^\tau\odif{t}\,(\dot\mu_x)^2
	\\
	 & \geq
	\label{eq:temp 2}
	\frac{\ab(\mu_x(\tau)-\mu_x(0))^2}{\tau},
\end{align}
where we used Eq.~\eqref{eq:mu_x}
and the Cauchy-Schwarz inequality
\begin{align}
	\label{eq:cs ineq}
	\ab(\int_0^\tau\odif{t}\,1^2)
	\ab(\int_0^\tau\odif{t}\,f(t)^2)
	\geq
	\ab(\int_0^\tau\odif{t}\,1\cdot f(t))^2
\end{align}
for an arbitrary function $f(t)$.
For the second to fourth terms
of Eq.~\eqref{eq:entropy production with normal},
we obtain
\begin{align}
	 & \quad
	\int_0^\tau\odif{t}\,
	\ab(
	\frac mT\Xi_{vv}
	+\frac Tm\frac{\Xi_{vv}}{\det\Xi}
	-2
	)
	\\
	 & =
	\int_0^\tau\odif{t}\,
	\ab[
		\frac mT\ab(\Xi_{vv}-\frac{\Xi_{xv}^2}{\Xi_{xx}})
		+\frac mT\frac{\Xi_{xv}^2}{\Xi_{xx}}
		+\frac1{\frac mT\ab(\Xi_{vv}-\frac{\Xi_{xv}^2}{\Xi_{xx}})}
		-2
	]        \\
	\label{eq:temp 3}
	 & \geq
	\int_0^\tau\odif{t}\,
	\frac mT\frac{\Xi_{xv}^2}{\Xi_{xx}}
	\\
	\label{eq:temp 4}
	 & =
	\int_0^\tau\odif{t}\,
	\frac mT\frac{\ab(\frac12\dot\Xi_{xx})^2}{\Xi_{xx}}
	\\
	 & =
	\int_0^\tau\odif{t}\,
	\frac mT\ab(\odv{}{t}\sqrt{\Xi_{xx}})^2
	\\
	\label{eq:temp 5}
	 & \geq
	\frac mT\frac{\ab(\sqrt{\Xi_{xx}(\tau)}-\sqrt{\Xi_{xx}(0)})^2}{\tau},
\end{align}
where we used
the inequality $ a+\frac{1}{a}\geq 2 $ for $ a > 0 $
in Eq.~\eqref{eq:temp 3}.
We also used
Eq.~\eqref{eq:xi_xx} in Eq.~\eqref{eq:temp 4},
and the Cauchy-Schwarz inequality Eq.~\eqref{eq:cs ineq}
in Eq.~\eqref{eq:temp 5}.
From Eqs.~\eqref{eq:entropy production with normal},
\eqref{eq:temp 2},
and \eqref{eq:temp 5}, we obtain
\begin{align}
	\label{eq:ineq given}
	\Sigma(\tau)
	\geq
	\frac{\gamma}{T}
	\frac{\ab(\mu_x(\tau)-\mu_x(0))^2+\ab(\sqrt{\Xi_{xx}(\tau)}-\sqrt{\Xi_{xx}(0)})^2}{\tau},
\end{align}
which yields the same form
as Eq.~\eqref{eq:known lower bound concretely},
consistently with the previous study \cite{dechant2019thermodynamicinterpretationwassersteindistance}.

However, the equality in Eq.~\eqref{eq:ineq given} is satisfied only when
$\mu_v$ and $\Xi$ are constant.
The condition that $\mu_v = \dot{\mu}_x$ is constant can be directly derived from the equality condition of Eq.~\eqref{eq:temp 2},
and the constancy of $\Xi$ is shown in Appendix.

\section{Minimization problem}

\subsection{Euler-Lagrange equations}
Using $k(t)$ and $r(t)$ as the control parameters,
we consider the time evolution paths
of the variables $\bm\mu(t)$ and $\Xi(t)$
satisfying the differential equations Eq.~\eqref{eq:dynamics}
as well as the initial and final conditions
$\bm{\mu}(0)=\bm{\mu}^{\mathrm{ini}},\bm{\mu}(\tau)=\bm{\mu}^{\mathrm{fin}}$,
and $\Xi(0)=\Xi^{\mathrm{ini}},\Xi(\tau)=\Xi^{\mathrm{fin}}$.
Among such paths,
we investigate what conditions should be imposed
on the path that minimizes the following quantity,
proportional to the entropy production $\Sigma(\tau)$
in Eq.~\eqref{eq:entropy production with normal}:
\begin{align}
	\label{eq:entropy production to minimize}
	\frac m\gamma\Sigma(\tau)
	 & =
	\int_0^\tau\odif{t}\,\ab(\frac{m}{T}\mu_v^2+\frac{m}{T}\Xi_{vv}+\frac{T}{m}\frac{\Xi_{xx}}{\det\Xi}-2).
\end{align}

To minimize Eq.~\eqref{eq:entropy production to minimize},
we define the Lagrangian
$\mathcal{L}(\bm\mu,\Xi,k,r,\{\lambda_i\},\dot{\bm\mu},\dot\Xi)$
by
\begin{align}
	\mathcal{L}
	:=
	\frac{m}{T}\mu_v^2+\frac{m}{T}\Xi_{vv}
	+\frac{T}{m}\frac{\Xi_{xx}}{\det\Xi}
	-2
	-\sum_{i=1}^5\lambda_if_i,
\end{align}
where $\{f_i\}$ represent the constraint terms arising
from the differential equations Eq.~\eqref{eq:dynamics}:
\begin{align}
	f_1 & := {\dot\mu}_x-\mu_v,                                                                   \\
	f_2 & := {\dot\mu}_v+\frac{k}{m}(\mu_x-r)+\frac{\gamma}{m}\mu_v,                              \\
	f_3 & := {\dot\Xi}_{xx}-2\Xi_{xv},                                                            \\
	f_4 & := {\dot\Xi}_{xv}+\frac{k}{m}\Xi_{xx}+\frac{\gamma}{m}\Xi_{xv}-\Xi_{vv},                \\
	f_5 & := {\dot\Xi}_{vv}+\frac{2k}{m}\Xi_{xv}+\frac{2\gamma}{m}\Xi_{vv}-\frac{2\gamma T}{m^2},
\end{align}
and $\{\lambda_i\}$ represent undetermined multipliers.

The optimal path must satisfy the Euler-Lagrange equation for each variable $\theta=\bm{\mu},\Xi, k,r$
\begin{align}
	\pdv{\mathcal L}\theta-\odv{}t\pdv{\mathcal L}{\dot\theta}=0.
\end{align}
The specific equations are as follows:
\begin{align}
	\label{eq:EL eq for mu_x}
	{\dot\lambda}_1-\frac{k}{m}\lambda_2                                                                                         & =0, \\
	\frac{2m}{T}\mu_v+\lambda_1+{\dot\lambda}_2-\frac{\gamma}{m}\lambda_2                                                        & =0, \\
	-\frac{T}{m}\frac{\Xi_{xv}^2}{(\det\Xi)^2}+{\dot\lambda}_3-\frac{k}{m}\lambda_4                                              & =0, \\
	-\frac{2T}{m}\frac{\Xi_{xx}\Xi_{xv}}{(\det\Xi)^2}+2\lambda_3+{\dot\lambda}_4-\frac{\gamma}{m}\lambda_4-\frac{2k}{m}\lambda_5 & =0, \\
	\frac{m}{T}-\frac{T}{m}\frac{\Xi_{xx}^2}{(\det\Xi)^2}+\lambda_4+{\dot\lambda}_5-\frac{2\gamma}{m}\lambda_5                   & =0, \\
	-\frac{1}{m}\lambda_2(\mu_x-r)-\frac{1}{m}\lambda_4\Xi_{xx}-\frac{2}{m}\lambda_5\Xi_{xv}                                     & =0, \\
	\label{eq:EL eq for k}
	\frac{k}{m}\lambda_2                                                                                                         & =0.
\end{align}

Assuming $k\neq 0$ (as setting $k=0$ would contradict controlling the system with a parabolic potential),
we can conclude that $\lambda_2=0$
and ${\dot\lambda}_1=0$ from Eqs.~\eqref{eq:EL eq for mu_x} and \eqref{eq:EL eq for k}.
Rearranging the remaining Euler-Lagrange equations yields the following:
\begin{align}
	\label{eq:mu_v const}
	\mu_v
	  & =\text{const.},                                                                                           \\
	\label{eq:lambda3}
	{\dot\lambda}_3
	  & =
	\frac{T}{m}\frac{\Xi_{xv}^2}{(\det\Xi)^2}+\frac{k}{m}\lambda_4,                                               \\
	{\dot\lambda}_4
	  & =
	-\frac{2T}{m}\frac{\Xi_{xx}\Xi_{xv}}{(\det\Xi)^2}-2\lambda_3+\frac{\gamma}{m}\lambda_4+\frac{2k}{m}\lambda_5, \\
	{\dot\lambda}_5
	  & =
	-\frac{m}{T}+\frac{T}{m}\frac{\Xi_{xx}^2}{(\det\Xi)^2}-\lambda_4+\frac{2\gamma}{m}\lambda_5,                  \\
	\label{eq:lambda balance}
	0 & =\lambda_4\Xi_{xx}+2\lambda_5\Xi_{xv}.
\end{align}

Observing these conditions,
it should be noted that the control parameter $r$ only appears
in Eq.~\eqref{eq:mu_v},
meaning that $r$ can be used to freely manipulate ${\dot\mu}_v$
if $k\neq 0$.
Thus,
$r$ can optimize $\bm{\mu}$ as seen in Sec.~\ref{subsec:opt_mu},
while $k$ can be used to optimize $\Xi$ as seen in Sec.~\ref{subsec:xi}.

\subsection{Optimization with respect to $\bm{\mu}$}
\label{subsec:opt_mu}
From Eqs.~(\ref{eq:dynamics}a),
(\ref{eq:dynamics}b),
and \eqref{eq:mu_v const},
we summarize the conditions for $\bm{\mu}$:
\begin{align}
	\label{eq:mu condition 1}
	{\dot\mu}_x
	 & =\mu_v,
	\\
	\label{eq:mu condition 2}
	{\dot\mu}_v
	 & =-\frac{k}{m}(\mu_x- r)-\frac{\gamma}{m}\mu_v,
	\\
	\label{eq:mu condition 3}
	\mu_v
	 & =\mathrm{const.}
\end{align}

As shown below,
when assuming that $\bm{\mu}$ is continuous,
it is not possible to set
$\bm{\mu}^\mathrm{ini}$ and $\bm{\mu}^\mathrm{fin}$ arbitrarily.
For example,
when setting $\mu_x^\mathrm{ini}$ and $\mu_x^\mathrm{fin}$ arbitrarily,
in order to satisfy the conditions Eqs.~\eqref{eq:mu condition 1}
and \eqref{eq:mu condition 3},
the path for $\mu_x$ and $\mu_v$ is determined as
\begin{align}
	\label{eq:opt mu x}
	\mu_x
	 & =
	\mu_x^\mathrm{ini}+\frac{\mu_x^\mathrm{fin}-\mu_x^\mathrm{ini}}{\tau} t,
	\\
	\label{eq:opt mu v}
	\mu_v
	 & =\frac{\mu_x^\mathrm{fin}-\mu_x^\mathrm{ini}}{\tau},
\end{align}
From Eq.~\eqref{eq:opt mu x},
we obtain
$\mu_v(0)=\mu_v(\tau)= (\mu_x^\mathrm{fin}-\mu_x^\mathrm{ini})/\tau$,
indicating that it is not possible
to specify both $\mu_v^\mathrm{ini}$
and $\mu_v^\mathrm{fin}$ arbitrarily.
Furthermore, the protocol for the control parameter $r$ is determined from
Eqs.~\eqref{eq:mu condition 2}, \eqref{eq:opt mu x}, and \eqref{eq:opt mu v}
as
\begin{align}
	r
	 & =
	\mu_x^\mathrm{ini}
	+\frac{\mu_x^\mathrm{fin}-\mu_x^\mathrm{ini}}{\tau}\ab( t+\frac{\gamma}{k}).
\end{align}

To see that Eqs.~\eqref{eq:opt mu x} and \eqref{eq:opt mu v}
are surely optimal, we note that
$\mu$ is included only in the form $\int_0^\tau\odif{t}\,\mu_v^2$
in the entropy production Eq.~\eqref{eq:entropy production to minimize}.
From Eq.~\eqref{eq:opt mu v}, we obtain
\begin{align}
	\label{eq:integral of square mu v}
	\int_0^\tau\odif{t}\,\mu_v^2
	 & =
	\frac{(\mu_x^\mathrm{fin}-\mu_x^\mathrm{ini})^2}{\tau}.
\end{align}
On the other hand, since we can see
\begin{align}
	\label{eq:integral of mu v}
	\int_0^\tau\odif{t}\,\mu_v
	=
	\mu_x^\mathrm{fin}-\mu_x^\mathrm{ini},
\end{align}
from Eq.~\eqref{eq:mu condition 1},
we obtain
\begin{align}
	\label{eq:ineq mu}
	\tau\int_0^\tau\odif{t}\,
	\mu_v^2\geq (\mu_x^\mathrm{fin}-\mu_x^\mathrm{ini})^2,
\end{align}
using Eq.~\eqref{eq:cs ineq} with $f(t)=\mu_v$.
Comparing Eqs.~\eqref{eq:integral of square mu v} and \eqref{eq:ineq mu},
we can confirm that
Eqs.~\eqref{eq:opt mu x} and \eqref{eq:opt mu v} are optimal.

If we allow noncontinuous behavior in $\bm{\mu}$ and $r$,
we can set
$\bm{\mu}^\mathrm{ini}$ and $\bm{\mu}^\mathrm{fin}$ arbitrarily,
which will be presented in Sec.~\ref{sec:discussion} together with
a similar discussion in the optimization with respect to $\Xi$.

\subsection{Optimization with respect to $\Xi$}
\label{subsec:xi}

From Eqs.~(\ref{eq:dynamics}c)-(\ref{eq:dynamics}e)
and \eqref{eq:lambda3}-\eqref{eq:lambda balance},
the conditions for $\Xi$ are summarized as follows:
\begin{subequations}\label{eq:xi_conditions}
	\begin{align}
		{\dot\Xi}_{xx}
		  & =2\Xi_{xv},              \\
		{\dot\Xi}_{xv}
		  & =
		-\frac{k}{m}\Xi_{xx}
		-\frac{\gamma}{m}\Xi_{xv}
		+\Xi_{vv},                   \\
		{\dot\Xi}_{vv}
		  & =
		-\frac{2k}{m}\Xi_{xv}
		-\frac{2\gamma}{m}\Xi_{vv}
		+\frac{2\gamma T}{m^2},      \\
		{\dot\lambda}_3
		  & =
		\frac{T}{m}\frac{\Xi_{xv}^2}{(\det\Xi)^2}
		+\frac{k}{m}\lambda_4,       \\
		{\dot\lambda}_4
		  & =
		-\frac{2T}{m}\frac{\Xi_{xx}\Xi_{xv}}{(\det\Xi)^2}
		-2\lambda_3
		+\frac{\gamma}{m}\lambda_4
		+\frac{2k}{m}\lambda_5,      \\
		{\dot\lambda}_5
		  & =
		-\frac{m}{T}+\frac{T}{m}\frac{\Xi_{xx}^2}{(\det\Xi)^2}
		-\lambda_4
		+\frac{2\gamma}{m}\lambda_5, \\
		0 & =
		\lambda_4\Xi_{xx}+2\lambda_5\Xi_{xv}.
		\label{eq:EL_k}
	\end{align}
\end{subequations}

Taking the time derivative of the condition Eq.~\eqref{eq:EL_k} gives
\begin{align}
	\notag
	0 & =
	\odv{}t\ab(\lambda_4\Xi_{xx}+2\lambda_5\Xi_{xv})  \\
	\notag
	  & =
	\lambda_4{\dot\Xi}_{xx}+2\lambda_5{\dot\Xi}_{xv}
	+{\dot\lambda}_4\Xi_{xx}+2{\dot\lambda}_5\Xi_{xv} \\
	\label{eq:EL_k_1}
	  & =
	-2\lambda_3\Xi_{xx}
	-\frac{2m}{T}\Xi_{xv}
	+2\lambda_5\Xi_{vv},
\end{align}
using Eqs.~(\ref{eq:xi_conditions}a), (\ref{eq:xi_conditions}b), (\ref{eq:xi_conditions}e), and (\ref{eq:xi_conditions}f).
Differentiating Eq.~\eqref{eq:EL_k_1}, we obtain
\begin{widetext}
	\begin{align}
		\label{eq:EL_k_2}
		0 & =
		\frac{m}{T}\ab(\frac{k}{m}\Xi_{xx}+\frac{\gamma}{m}\Xi_{xv}-2\Xi_{vv})
		+\frac{T}{m}\frac{\Xi_{xx}}{\det\Xi}
		-2\lambda_3\Xi_{xv}-\lambda_4\Xi_{vv}+\frac{2\gamma T}{m^2}\lambda_5,
	\end{align}
	and, differentiating the above further,
	we obtain the evolution equation of the optimal $k$ as
	\begin{align}
		\dot{k}
		=
		-4k\frac{\Xi_{xv}}{\Xi_{xx}}
		-\frac{T}{\Xi_{xx}}
		\ab[\frac{4T}{m}\frac{\Xi_{xv}}{\det\Xi}
			+\frac{3\gamma}{m}\ab(
			\frac{m}{T}\Xi_{vv}
			+\frac{T}{m}\frac{\Xi_{xx}}{\det\Xi}
			-2
			)
			+\frac{2\gamma T}{m^2}\ab(-2\lambda_4+\frac{3\gamma}{m}\lambda_5)
		].
		\label{eq:dotk}
	\end{align}
\end{widetext}
Consequently, satisfying Eq.~\eqref{eq:EL_k} over the interval $t\in (0,\tau)$
is equivalent with satisfying Eq.~\eqref{eq:dotk} for $t\in (0,\tau)$ along with the conditions in Eqs.~\eqref{eq:EL_k},
\eqref{eq:EL_k_1}, and \eqref{eq:EL_k_2} at $t=0$.

We here note that
choosing arbitrary $\Xi^{\mathrm{ini}}$ and $\Xi^{\mathrm{fin}}$
may be impossible,
because the optimal path is uniquely determined once $\Xi(0)$ and $k(0)$ are specified.
Specifically, it is determined as follows.

When $\Xi(0)$ and $k(0)$ are specified,
the initial values $\lambda_3(0),\lambda_4(0),\lambda_5(0)$
are determined from Eqs.~\eqref{eq:EL_k}, \eqref{eq:EL_k_1},
and \eqref{eq:EL_k_2}.
Thus, using Eqs.~(\ref{eq:xi_conditions}a)-(\ref{eq:xi_conditions}f) and \eqref{eq:dotk},
all variable values
$\Xi(t)$, $\lambda_3(t)$, $\lambda_4(t)$, $\lambda_5(t)$,
and $k(t)$ at $t > 0$ are determined.
Hence, $\Xi(\tau)$ is also determined,
which means that
we cannot specify $\Xi^\mathrm{fin}$ arbitrarily.

\section{Numerical simulation}
In this section,
we perform numerical simulations of Eq.~\eqref{eq:xi_conditions}, which are related
to only $\Xi$ not $\bm\mu$, to confirm
the findings
in Sec.~\ref{subsec:xi}.
Specifically,
we verify the constraint of the specification of $\Xi^\mathrm{fin}$
and whether the conditions for the entropy production
to reach its lower bound are indeed satisfied by our simulation results.

When $\Xi^\mathrm{ini}$ and $k^\mathrm{ini}$ are given,
we refer to the time evolution of the variables determined by
Eqs.~\eqref{eq:xi_conditions}-\eqref{eq:dotk}
as the optimal path, and we perform numerical simulations on it.

In the simulations shown below,
we set $m=\gamma=T=\tau=1$ as an example,
but similar results can be obtained with other settings.

\subsection{Possible final covariance matrices $\Xi^\mathrm{fin}$}
\label{subsec:reachable}
\begin{figure}[htbp]
	\centering
	\includegraphics[width=\columnwidth]{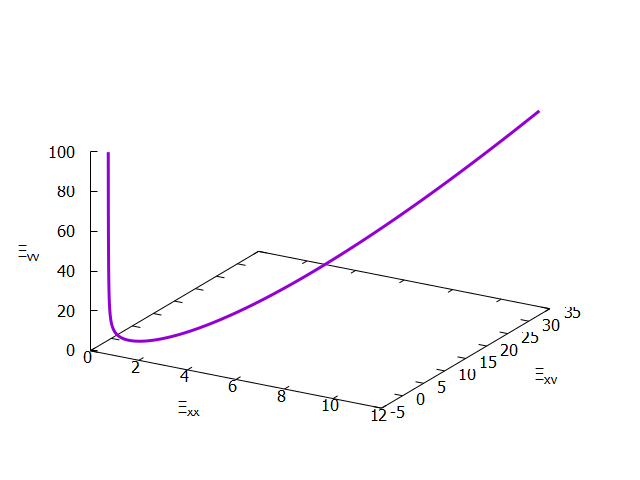}
	\caption{
		The set $[\Xi_{xx}(\tau),\Xi_{xv}(\tau),\Xi_{vv}(\tau)]$
		when $\Xi^\mathrm{ini}$ is fixed and $k^\mathrm{ini}$ is varied
		from -0.344 to 1.569 incrementing by 0.001.
		We exclude the case of
		$k^\mathrm{ini}<-0.344$ and $k^\mathrm{ini}>1.569$
		to prevent the drawing range from becoming excessively large.
	}
	\label{fig:goals}
\end{figure}

We set $\Xi^\mathrm{ini}_{xx}=1$, $\Xi^\mathrm{ini}_{xv}=0$, $\Xi^\mathrm{ini}_{vv}=1$ as the initial condition of $\Xi$,
and calculate $\Xi(\tau)$ for various values of $k^\mathrm{ini}$,
and plot the resulting set $[\Xi_{xx}(\tau),\Xi_{xv}(\tau),\Xi_{vv}(\tau)]$
(Fig.~\ref{fig:goals}).

Figure~\ref{fig:goals} illustrates
the reachable $\Xi(\tau)$ of the optimal path from $\Xi^\mathrm{ini}$,
in other words,
$\Xi^\mathrm{fin}$ for which the continuous optimal path
from $\Xi^\mathrm{ini}$ exists.
Therefore,
the continuous optimal path does not necessarily exist for
arbitrary $\Xi^\mathrm{fin}$.

\subsection{Comparison of optimal and nonoptimal paths}
Here we aim to verify that
the entropy production along the optimal path,
determined by Eqs.~\eqref{eq:xi_conditions}-\eqref{eq:dotk},
is smaller than that along the nonoptimal path,
which evolves according to Eqs.~(\ref{eq:dynamics}c)-(\ref{eq:dynamics}e),
choosing $k(t)$
that is specially manipulated so as to share
$\Xi^\mathrm{ini}$ and $\Xi^\mathrm{fin}$ with
the optimal path.

As mentioned in Sec.~\ref{subsec:xi},
it is not possible to find the optimal path
for arbitrary $\Xi^\mathrm{ini}$ and $\Xi^\mathrm{fin}$.
Therefore we use brute-force approach to find the optimal
and nonoptimal paths with the same $\Xi^\mathrm{ini}$ and $\Xi^\mathrm{fin}$.
Concretely,
we search for a pair consisting of an optimal path and a nonoptimal path
with close values of $\Xi(\tau)$
\footnote{
	We evaluate the difference in $\Xi(\tau)$ for the two paths
	using the Euclidean distance in the coordinates $(\Xi_{xx},\Xi_{xv},\Xi_{vv})$.
} through the following steps:
\begin{enumerate}
	\item
	      \label{item:det ini}
	      Arbitrarily determine $\Xi^\mathrm{ini}$.
	\item
	      \label{item:calc opt path}
	      Arbitrarily determine $k^\mathrm{ini}$,
	      and calculate an optimal path starting
	      from $\Xi^\mathrm{ini}$.
	\item
	      \label{item:make opt path set}
	      Create a set of optimal paths by repeating step (ii) with various values of $k^\mathrm{ini}$.
	\item
	      \label{item:calc non-opt path}
	      Set an arbitrary time-independent $k$
	      and calculate a nonoptimal path that starts
	      from $\Xi^\mathrm{ini}$.
	      (Note that the use of
	      time-independent $k$ is for simplicity and this $k$
	      may not give $\Xi^\mathrm{fin}$ of the optimal path.)
	\item
	      \label{item:make pair}
	      From the set of optimal paths created
	      in step (iii),
	      select the optimal path with $\Xi(\tau)$ closest
	      to that of the nonoptimal path calculated
	      in step (iv) and pair these paths.
	\item
	      \label{item:search pair}
	      Repeat steps (iv) and (v)
	      changing the value of $k$,
	      to search for a pair with especially close values of $\Xi(\tau)$.
	\item
	      Repeat steps (i)-(vi)
	      changing the value of $\Xi^\mathrm{ini}$,
	      to find the pair with the closest value of $\Xi(\tau)$.
\end{enumerate}
We present below the paths that constitute the pair found in this way.

\begin{figure}[htbp]
	\centering
	\includegraphics[width=\columnwidth]{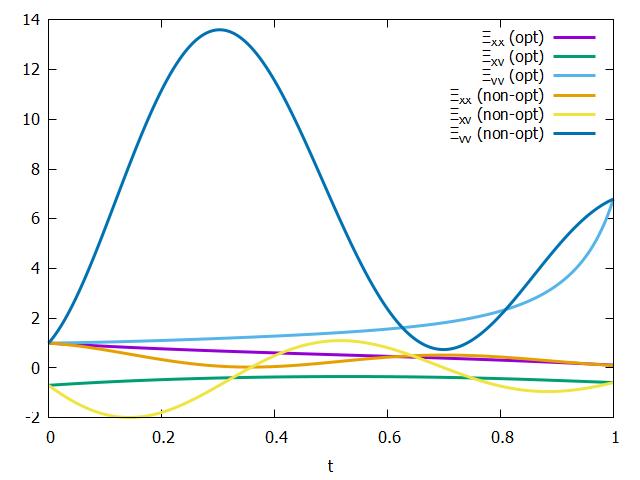}
	\caption{
		The time evolution of $\Xi$
		for the optimal path and the nonoptimal path.
		For example,
		$\Xi_{xx}$ for the optimal path (purple line)
		and $\Xi_{xx}$ for the nonoptimal path (orange line)
		exhibit close values at $t=\tau(=1)$.
	}
	\label{fig:xi}
\end{figure}

When $\Xi_{xx}^\mathrm{ini}=1$, $\Xi_{xv}^\mathrm{ini}=-0.698$
and $\Xi_{vv}^\mathrm{ini}=1$,
we found that
the optimal path with $k^\mathrm{ini}=0.187$
and the nonoptimal path with time-independent $k(t)=18.235$ evolve
as shown in Fig.~\ref{fig:xi}.
From Fig.~\ref{fig:xi}, we can see that $\Xi(\tau)$ of both paths
are almost identical.

We denote the terms
in the entropy production rate Eq.~\eqref{eq:entropy production rate}
related to $\Xi$ by $\sigma_\Xi$ as
\begin{align}
	{\sigma}_\Xi(t)
	 & :=
	\frac{m}{T}\Xi_{vv}+\frac{T}{m}\frac{\Xi_{xx}}{\det\Xi}-2,
\end{align}
and its integral,
which we aim to minimize here,
by $\Sigma_\Xi$ as
\begin{align}
	\label{eq:Sigma Xi}
	{\Sigma}_\Xi(\tau)
	 & :=
	\int_0^\tau\odif{t}\,{\sigma}_\Xi(t).
\end{align}
The lower bound shown in the previous studies
\cite{dechant2019thermodynamicinterpretationwassersteindistance,nakazato2021} corresponds to
\begin{align}
	\label{eq:lower bound comperison}
	\Sigma_\Xi(\tau)\geq
	\frac mT\frac{\ab(\sqrt{\Xi_{xx}(\tau)}-\sqrt{\Xi_{xx}(0)})^2}\tau,
\end{align}
as indicated in Eq.~\eqref{eq:temp 5}.

\begin{figure}[htbp]
	\centering
	\includegraphics[width=\columnwidth]{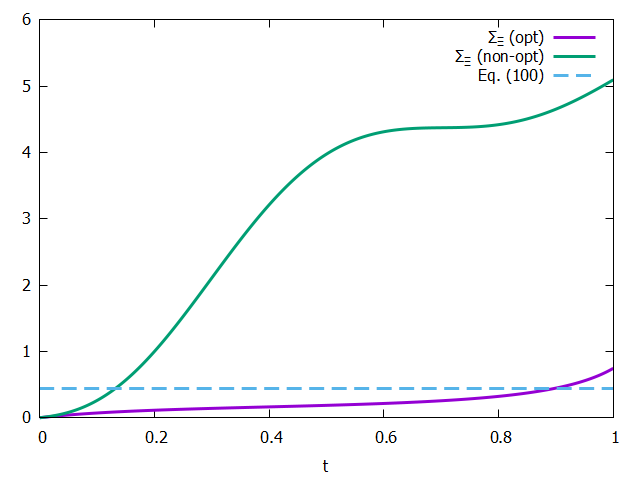}
	\caption{
		The entropy production
		$\Sigma_\Xi(t)$
		in Eq.~\eqref{eq:Sigma Xi}
		for the optimal and nonoptimal paths,
		and the lower bound Eq.~\eqref{eq:lower bound comperison}
		shown in the previous studies.
	}
	\label{fig:sigma}
\end{figure}

From Fig.~\ref{fig:sigma},
we can confirm that
the entropy production $\Sigma_\Xi(\tau)$ in Eq.~\eqref{eq:Sigma Xi}
of the optimal path is smaller than that of the nonoptimal path.
We can also confirm that
the entropy production $\Sigma_\Xi(\tau)$ in both paths
is greater than the known lower bound
in Eq.~\eqref{eq:lower bound comperison}.
This is because, as mentioned in Sec.~\ref{subsec:nd},
the lower bound in Eq.~\eqref{eq:lower bound comperison} cannot be achieved when $\Xi$ is not constant.

\section{Discussion}\label{sec:discussion}
As described in Sec.\ref{subsec:opt_mu} and Sec.\ref{subsec:xi},
we found that the degrees of freedom
for setting the initial and final conditions are limited.
However,
if the control parameters $k$ and $r$
are allowed to exhibit noncontinuous behavior,
the degrees of freedom for setting the conditions can increase.

We consider $\mu_v^*$,
where the endpoints $\mu_v(0)$ and $\mu_v(\tau)$ of $\mu_v(t)$
in Eq.~\eqref{eq:opt mu v} are changed as
\begin{align}
	\label{eq:mu v star}
	\mu_v^*
	=
	\begin{cases}
		\mu_v^\mathrm{ini} & t=0          \\
		\mu_v^\mathrm{mid} & t\in(0,\tau) \\
		\mu_v^\mathrm{fin} & t=\tau
	\end{cases},
\end{align}
using $\mu_v^\mathrm{mid}:={(\mu_x^\mathrm{fin}-\mu_x^\mathrm{ini})}/{\tau}$ that is the same as Eq.~\eqref{eq:opt mu v}.
Equation \eqref{eq:mu v star} satisfies the initial and final conditions of $\mu_v$,
while Eq.~\eqref{eq:opt mu v} did not satisfy them.
On the other hand, similar to Eq.~\eqref{eq:opt mu v},
Eq.~\eqref{eq:mu v star} minimizes the entropy production because $\int\odif{t}\,\mu_v(=\mu_x)$ and $\int_0^\tau\odif{t}\,\mu_v^2$ remain unchanged from Eqs.~\eqref{eq:opt mu x} and \eqref{eq:integral of square mu v},
as the endpoints do not contribute to the integral.

The corresponding $r$ that realizes $\mu_v^*$ in Eq.~\eqref{eq:mu v star}
is derived from Eqs.~\eqref{eq:mu condition 2},
\eqref{eq:opt mu x},
and \eqref{eq:mu v star} as
\begin{align}
	r
	=
	\mu_x^\mathrm{ini}+\mu_v^\mathrm{mid} t
	+\frac{\gamma}{k}\mu_v^*
	+\frac{m}{k} {\dot\mu}_v^*.
\end{align}
Here,
from Eq.~\eqref{eq:mu v star},
${\dot\mu}_v^*$ is given by
\begin{align}
	{\dot\mu}_v^*
	=
	-2\delta(t)(\mu_v^\mathrm{ini}-\mu_v^\mathrm{mid})
	+2\delta(t-\tau)(\mu_v^\mathrm{fin}-\mu_v^\mathrm{mid}),
\end{align}
where $\delta$ is the Dirac delta function,
which satisfies the following property for integration:
\begin{align}
	\int_{-\infty}^t\mathrm{d}t'\,\delta(t')
	=
	\begin{cases}
		0           & t < 0 \\
		\frac{1}{2} & t=0   \\
		1           & t > 0
	\end{cases}.
\end{align}

Similarly,
we can apply the above discussion for $\mu_v$
to the continuous optimal path $(\Xi, k)$.
Instead of $(\Xi, k)$,
we consider the path $(\Xi^*, k^*)$,
where, using constants $c_0$ and $c_\tau$,
$k^*$ is defined by
\begin{align}
	\label{eq:k star}
	k^*(t)=k+c_0\delta(t)+c_\tau\delta(t-\tau),
\end{align}
which differs from $k$ only at $t=0$ and $t=\tau$.
$\Xi^*$ is defined to take the same values as $\Xi$
for $t\in (0,\tau)$,
but to take noncontinuous values at $t=0$ and $t=\tau$
as follows:
\begin{subequations}\label{eq:xi star}
	\begin{align}
		\Xi_{xx}(0)      & =\Xi_{xx}^*(0),                                                                     \\
		\Xi_{xv}(0)      & =\Xi_{xv}^*(0)-\frac{c_0}{2m}\Xi_{xx}^*(0),                                         \\
		\Xi_{vv}(0)      & =\Xi_{vv}^*(0)-\frac{c_0}{m}\Xi_{xv}^*(0),                                          \\
		\Xi_{xx}^*(\tau) & =\Xi_{xx}(\tau),                                                                    \\
		\Xi_{xv}^*(\tau) & =\Xi_{xv}(\tau)-\frac{c_\tau}{2m}\Xi_{xx}(\tau),                                    \\
		\Xi_{vv}^*(\tau) & =\Xi_{vv}(\tau)-\frac{c_\tau}{m}\Xi_{xv}(\tau)+\frac{c_\tau^2}{2m^2}\Xi_{xx}(\tau).
	\end{align}
\end{subequations}

These results can be derived from the integration of
\begin{subequations}\label{eq:xi star condition}
	\begin{align}
		{\dot\Xi}^*_{xx}
		 & =2\Xi^*_{xv}, \\
		{\dot\Xi}^*_{xv}
		 & =
		-\frac{k}{m}\Xi^*_{xx}
		-\frac{\gamma}{m}\Xi^*_{xv}
		+\Xi^*_{vv},     \\
		{\dot\Xi}^*_{vv}
		 & =
		-\frac{2k}{m}\Xi^*_{xv}
		-\frac{2\gamma}{m}\Xi^*_{vv}
		+\frac{2\gamma T}{m^2},
	\end{align}
\end{subequations}
corresponding to Eqs.~(\ref{eq:xi_conditions}a)-(\ref{eq:xi_conditions}c),
in the vicinity of $t=0$ and $t=\tau$.
Below we show an example of such an integration.

Considering the integral of ${\dot\Xi}_{xv}^*$
in the region near $t=0$,
we obtain
\begin{align}
	\int_0^\varepsilon\odif{t}\,{\dot\Xi}_{xv}^*
	 & =\Xi_{xv}^*(\varepsilon)-\Xi_{xv}^*(0)                \\
	 & =\Xi_{xv}(\varepsilon)-\Xi_{xv}^*(0)                  \\
	\label{eq:step}
	 & \to\Xi_{xv}(0)-\Xi_{xv}^*(0)\quad (\varepsilon\to 0),
\end{align}
denoting a small positive number by $\varepsilon$.
On the other hand,
using Eqs.~\eqref{eq:k star} and (\ref{eq:xi star condition}b),
we obtain
\begin{align}
	\int_0^\varepsilon\odif{t}\,{\dot\Xi}_{xv}^*
	 & =
	\int_0^\varepsilon\odif{t}\,\ab(-\frac{k^*}{m}\Xi^*_{xx}
	-\frac{\gamma}{m}\Xi^*_{xv}
	+\Xi^*_{vv})
	\\
	 & =
	\int_0^\varepsilon\odif{t}\,\ab(-\frac{k}{m}\Xi^*_{xx}
	-\frac{\gamma}{m}\Xi^*_{xv}
	+\Xi^*_{vv}
	-\frac{c_0}{m}\delta(t)\Xi^*_{xx}
	)
	\\
	 & =
	\int_0^\varepsilon\odif{t}\,\ab(-\frac{k}{m}\Xi^*_{xx}
	-\frac{\gamma}{m}\Xi^*_{xv}
	+\Xi^*_{vv}
	)
	-\frac{c_0}{2m}\Xi^*_{xx}(0)
	\\
	\label{eq:jump}
	 & \to-\frac{c_0}{2m}\Xi_{xx}^*(0)\quad (\varepsilon\to 0).
\end{align}
Comparing Eq.~\eqref{eq:step} with Eq.~\eqref{eq:jump},
we have
\begin{align}
	\Xi_{xv}(0)-\Xi_{xv}^*(0)=-\frac{c_0}{2m}\Xi_{xx}^*(0),
\end{align}
which leads to Eq.~(\ref{eq:xi star}b).

As shown in Fig.~\ref{fig:goals},
the reachable $\Xi(\tau)$ of the continuous optimal path
from $\Xi^\mathrm{ini}$ forms a one-dimensional curve
because we can only control $k^\mathrm{ini}$.
However,
if we allow such noncontinuous changes as Eq.~\eqref{eq:k star},
we become able to control $c_0$ and $c_\tau$,
which may be regarded as two additional degrees of freedom
of the control parameters.
Therefore,
the reachable $\Xi^*(\tau)$ may potentially make a nonzero-volume region
in three-dimensional variable space.
As a result,
allowing the noncontinuous $k(t)$
may enable us to find the optimal path for certain $\Xi^\mathrm{ini}$
and $\Xi^\mathrm{fin}$ that were unattainable with the continuous $k(t)$.
For example,
if we determine $c_\tau \neq 0$
and perform a jump from an arbitrary point in Fig.~1
according to Eqs.~(\ref{eq:xi star}d)-(\ref{eq:xi star}f),
$\Xi_{xx}$ remains unchanged while $\Xi_{xv}$ changes.
Therefore, we can arrive at a point that is not depicted in Fig.~1.

\section{Conclusion}
\label{sec:conclusion}
In this study,
we have explored the minimization of the entropy production
in a one-dimensional underdamped Langevin system
constrained by a time-dependent parabolic potential,
focusing specifically on the processes
where the probability distribution is restricted to be normal.
By considering the mean and covariance matrix of the system,
we derived the necessary conditions
that must be satisfied to achieve the lower bound of the entropy production.
These conditions were obtained by formulating
and solving the Euler-Lagrange equations
associated with the minimization problem,
where the stiffness $k(t)$ and the center $r(t)$ of the potential
are treated as control parameters.

Our analysis
and numerical simulation revealed that,
unlike in the overdamped systems \cite{nakazato2021},
there are inherent limitations
in arbitrarily specifying the initial
and final distributions $p^{\mathrm{ini}}$ and $p^{\mathrm{fin}}$
in the underdamped case.
These limitations arise from the limited degrees of freedom
of our control parameters.

Moreover,
we compared an optimal path with a nonoptimal path
through the numerical simulation
and showed that the optimal control indeed results
in lower entropy production,
validating the effectiveness of the derived conditions.

In addition,
we considered the possibility of increasing the degrees of freedom
in specifying the initial and final distributions
by allowing the control parameters to exhibit noncontinuous behavior,
such as incorporating discontinuities or
Dirac delta functions in $k(t)$ and $r(t)$.
This approach can, in theory,
enable the system to achieve a wider range of initial and final conditions.
However,
such control strategies may not be practical or physically realizable
due to the instantaneous changes required in the control parameters.

Our findings extend the existing knowledge
of the entropy production minimization from overdamped
to underdamped Langevin systems,
highlighting the additional complexities introduced by inertia.
The results emphasize the importance of
considering the system dynamics
and control limitations when designing protocols
for minimizing energy dissipation in stochastic thermodynamic systems.

Future work could focus on exploring alternative control strategies
that offer a greater flexibility in specifying the initial and final conditions,
such as introducing time-dependent temperature profiles
or additional control forces.

\section*{Acknowledgments}
This work was supported by JST, the establishment of university fellowships towards the creation of science technology innovation, Grant No. JPMJFS2101.

\appendix
\renewcommand{\theequation}{A\arabic{equation}}
\setcounter{equation}{0}
\section*{APPENDIX: CONSTANCY OF $\Xi$ WHEN THE KNOWN LOWER BOUND IS ACHIEVED}
\label{sec:app1}
For the equality in Eq.~\eqref{eq:ineq given},
we need the equalities in Eqs.~\eqref{eq:temp 3} and \eqref{eq:temp 5}.
The condition for the equality in Eq.~\eqref{eq:temp 5}
is given by $\odv{}t\sqrt{\Xi_{xx}} = \mathrm{const.}$,
so $\Xi_{xx}$ can be expressed using constants $a$ and $b$ as
\begin{align}
	\label{eq:xx ab}
	\Xi_{xx} & = (at+b)^2,
\end{align}
and $\Xi_{xv}$ is also obtained as
\begin{align}
	\label{eq:xv ab}
	\Xi_{xv} & = a(at+b)
\end{align}
from Eq.~\eqref{eq:xi_xx}.
Since the equality condition in Eq.\eqref{eq:temp 3} is given by
\begin{align}
	\frac mT\ab(\Xi_{vv} - \frac{\Xi_{xv}^2}{\Xi_{xx}}) = 1,
\end{align}
we have
\begin{align}
	\label{eq:vv ab}
	\Xi_{vv} & = a^2 + \frac{T}{m}.
\end{align}
Substituting Eqs.~\eqref{eq:xx ab}, \eqref{eq:xv ab}, and \eqref{eq:vv ab} into Eq.~\eqref{eq:xi_vv}, we obtain
\begin{align}
	ka(at+b) = -\gamma a^2.
\end{align}
Thus, it is necessary that $a=0$ or
\begin{align}
	\label{eq:k ab}
	k = -\frac{\gamma a}{at+b}.
\end{align}
Since Eq.~\eqref{eq:xi_xv} is inconsistent with Eqs.~\eqref{eq:xx ab}, \eqref{eq:xv ab}, \eqref{eq:vv ab}, and \eqref{eq:k ab},
$a=0$ must hold.
In this case,
from Eqs.~\eqref{eq:xx ab}, \eqref{eq:xv ab}, \eqref{eq:vv ab}, and \eqref{eq:xi_xv},
we have
\begin{align}
	\Xi_{xx} & = \frac{T}{k} = \mathrm{const.},
	\\
	\Xi_{xv} & = 0,
	\\
	\Xi_{vv} & = \frac{T}{m}.
\end{align}

\end{document}